\documentclass[12pt]{iopart}

\usepackage{iopams}  
\begin{document}

\title{A 5-Dimensional Spherical Symmetric Solution in Einstein-Yang-Mills Theory With Gauss-Bonnet Term}

\author{R J Slagter}

\address{Institute of Physics, University of Amsterdam \\ and \\ ASFYON, Astronomisch Fysisch Onderzoek Nederland, 1405EP Bussum, 
The Netherlands }
\ead{info@asfyon.nl}
\begin{abstract}
We present a numerical solution on a 5-dimensional spherically symmetric space time, in Einstein-Yang-Mills-Gauss-Bonnet theory using a
two point boundary value routine.
It turns out that the Gauss-Bonnet contribution has a profound influence on the behaviour of the particle-like solution: 
it increases the number of nodes of the YM field. When a negative cosmological constant in incorporated in the model, it turns out that 
there is no horizon and no singular behaviour of the model. For positive cosmological constant the model has singular behaviour.
\end{abstract}

\maketitle
\section{Introduction}
In recent years higher dimensional gravity is attracting much interest. One reason is the possibility that these higher
dimensions could be detectable at CERN.
The possibility that space time may have more than four dimensions is initiated by high energy physics
and inspired by D-brane ideology in string theory.
Our 4-dimensional space time (brane) is embedded in the 5-dimensional bulk. It is assumed that all the standard model
degrees of freedom reside on the brane, where as gravity can propagate into the bulk \cite{Ran}.
The effect of string theory on classical gravitational physics is investigated by the low-energy effective action. 
If our 5-dimensional space time is obtained as an effective theory, the matter fields, for example the U(1) field, can exists
in the bulk. 
In General Relativity(GR), gravitating non-Abelian gauge field, i.e., the Yang-Mills(YM) field, can be regarded as the most natural
generalization of Einstein-Maxwell(EM) theory. In particular, particle-like, soliton-like  and black hole solutions in the combined 
Einstein-Yang-Mills(EYM) models, shed new light on the complex features of compact object in these models. See \cite{Vol} for an overview.

The reason for adding a cosmological constant to these models, was inspired by the study of the so-called AdS/CFT correspondence \cite{Mal,Hos},
since the 5-dimensional Einstein gravity with cosmological constant gives a description of 4-dimensional conformal field theory
in large N limit. 
Brane world scenarios predict a negative cosmological constant. 
There is a relationship between the FRW equations controlling the cosmological expansion and the formulas that relate
the energy and entropy of the CFT \cite{Ver}, indicating that both sets of equations may have a common origin.
In the AdS-brane cosmological models, the AdS/CFT model describes a CFT dominated universe as a co-dimension one brane, with fixed tension,
in the background of an AdS black hole \cite{Gub}. The brane starts out inside the black hole,  passes through the  horizon and keeps expanding
until it reaches a maximal radius, after which it contracts and falls back into the black hole. At these moments of horizon crossing,
it turns out that the FRW equation turns into an equation that expresses the entropy density in terms of the energy density and coincides
with  the entropy of the CFT.
However, in these models, one adds on an artificial way tension into the equations. More general, one could solve the equations of Einstein together
with the matter field equations, for example, the YM field and try to obtain the same correspondence.

String theory also predicts quantum corrections to classical gravity theory and the Gauss-Bonnet(GB) term is the only one leading
to second order differential equations in the metric. 
In the 4-dimensional EYM-GB model with a dilaton field (EYMD-GB) \cite{Don,Tor}, it was found that the GB contribution 
can lead to possible new types of dilatonic black holes. Further, for a critical GB coupling $\kappa >\kappa_{cr}$ the solutions cease to
exist.
The AdS/CFT correspondence can also be investigated in the Einstein-GB gravity. For a recent overview, see \cite{Og}.  
From the viewpoint of AdS/CFT correspondence, it is argued that the GB term in the bulk corresponds to the next leading order 
corrections in the $\frac{1}{N}$ expansion of a CFT.  Further, it is argued that the entropy of an Einstein-GB black hole and the CFT
entropy induced on the brane are equal in the high temperature limit.

In this paper we investigate the possibility of regular and singular solutions in the 5-dimensional EYM-GB  model and the effect of 
a cosmological constant on the behaviour of the solutions. 
\section{The model}
The action of the model under consideration is \cite{Okuy}
\begin{equation}
{\cal S}=\frac{1}{16\pi}\int d^5x\sqrt{-g_5}\Bigl[\frac{1}{ G_5}(R-\Lambda)+\kappa(R_{\mu\nu\alpha\beta}
R^{\mu\nu\alpha\beta}-4R_{\alpha\beta}R^{\alpha\beta}+R^2)-\frac{1}{g^2}Tr{\bf F^2}\Bigr],
\end{equation}
with $G_5$ the gravitational constant, $\Lambda$ the cosmological constant, $\kappa$ the Gauss-Bonnet
coupling and $g$ the gauge coupling.
The coupled set of equations of the EYM-GB system will then become
\begin{eqnarray}
\Lambda  g_{\mu\nu}+G_{\mu\nu}-\kappa GB_{\mu\nu}=8\pi G_5 T_{\mu\nu}, 
\end{eqnarray}
\begin{eqnarray}
{\cal D}_\mu F^{\mu\nu a}=0,
\end{eqnarray}
with the Einstein tensor
\begin{eqnarray}
G_{\mu\nu}= R_{\mu\nu}-\frac{1}{2}g_{\mu\nu} R, 
\end{eqnarray}
and Gauss-Bonnet tensor  
\begin{eqnarray}
GB_{\mu\nu}=\frac{1}{2}g_{\mu\nu}\Bigl( R_{\gamma\delta\lambda\sigma}R^{\gamma\delta\lambda\sigma}
-4R_{\gamma \delta}R^{\gamma\delta} +R^2 \Bigr) -2RR_{\mu\nu}+4R_{\mu\gamma}{R^{\gamma}}_{\nu} \cr +
4R_{\gamma\delta}{{{R^{\gamma}}_{\mu}}^{\delta}}_{\nu}
 -2R_{\mu\gamma\delta\lambda}{R_{\nu}}^{\gamma\delta\lambda},
\end{eqnarray}
Further, with 
$R_{\mu\nu}$ the Ricci tensor and $T_{\mu\nu}$ the energy-momentum tensor
\begin{eqnarray}
T_{\mu\nu}={\bf Tr}F_{\mu\lambda}F_\nu^\lambda -\frac{1}{2}g_{\mu\nu}{\bf Tr}F_{\alpha\beta}F^{\alpha\beta},
\end{eqnarray}
and with $F_{\mu\nu}^a=\partial_\mu A_\nu^a -\partial_\nu A_\mu^a +g\epsilon^{abc}A_\mu^b A_\nu^c $, and 
${\cal D}_\alpha F_{\mu\nu}^a=\nabla_\alpha F_{\mu\nu}^a+g\epsilon^{abc}A_\alpha^b F_{\mu\nu}^c$
where $A_\mu^a$ represents  the YM potential.

Consider now the spherically symmetric 5-dimensional space time
\begin{equation}
ds^2=-\frac{F}{E^2}dt^2+\frac{1}{F}dr^2+r^2\Bigl(d\psi^2+\sin^2\psi(d\theta^2+\sin^2\theta^2d\varphi^2)\Bigr),
\end{equation}
with the YM parameterization
\begin{eqnarray}
A_t^{(a)}=A_r^{(a)}=0, A_\psi^{(a)}=\Bigl(0,0,W\Bigr),\cr
A_\theta^{(a)}=\Bigl(W\sin\psi,-\cos\psi ,0\Bigr),\cr
A_\varphi^{(a)}=\sin\theta\Bigl(\cos\psi,W\sin\psi,\frac{-1}{\tan\theta} \Bigr),
\end{eqnarray}
where $F$ and $W$ are functions of and $t$ and $r$. It turns out that no time evolution of the metric component $E$ can be
found from the equations, so $E$ depends only on r.
 The equations become (we take $g=1$)

\begin{equation}
F_r=\frac{2r(1-F)-\frac{2}{3}\Lambda r^3-\frac{G_5}{r}(1-W^2)^2-G_5r(FW_r^2+\frac{E^2}{F}W_t^2)}{r^2+4\kappa (1-F)},
\end{equation}
\begin{equation}
E_r=\frac{-G_5 rE(F^2W_r^2+E^2W_t^2)}{F^2(r^2+4\kappa(1-F))},
\end{equation}
\begin{equation}
F_t=\frac{-2G_5rFW_tW_r}{r^2+4\kappa(1-F)}
\end{equation}
and 
\begin{equation}
W_{tt}=\frac{F^2}{E^2}W_{rr}+\frac{W_tF_t}{F}+\frac{F^2W_r}{E^2}\Bigl(\frac{-E_r}{E}+\frac{F_r}{F}+\frac{1}{r}\Bigr)-\frac{2WF(W^2-1)}{E^2r^2}.
\end{equation}

\section{Numerical solutions}

The independent field equations  then read
\begin{equation}
W''=W'\Bigl(\frac{2r(F-1)+\frac{2}{3}\Lambda r^3+\frac{G_5}{r}(1-W^2)^2}{F(r^2+4\kappa (1-F))}-\frac{1}{r}\Bigr)+\frac{2W(W^2-1)}{Fr^2},
\end{equation}
\begin{equation}
F'=\frac{2r(1-F)-\frac{2}{3}\Lambda r^3-\frac{G_5}{r}(1-W^2)^2-G_5rFW'^2}{r^2+4\kappa (1-F)},
\end{equation}
while the equation for $E$ decouples and can be integrated:
\begin{equation}
E=e^{-G_5\int\frac{rW'^2}{r^2+4\kappa (1-F)}dr}.
\end{equation}
The equations are easily solved with an ODE solver and checked with MAPLE. 
We will take for the initial value of $W$ the usual form $W(0)=1-br^2$. Then the
other variables can be expanded around $r=0$:
\begin{equation}
E(0)= a+\frac{G_5ab^2}{4\kappa(c-1)}r^4 + ...\quad F(0)=c+\frac{1}{4\kappa}r^2 + ....
\end{equation}
So we have 3 initial parameters and 4 fundamental constants $\Lambda, G_5$, $g$ and $\kappa$.

We solved the equations with a two point boundary value solver.

\input epsf
\centerline{
\hfill\epsfysize=45mm\epsfbox{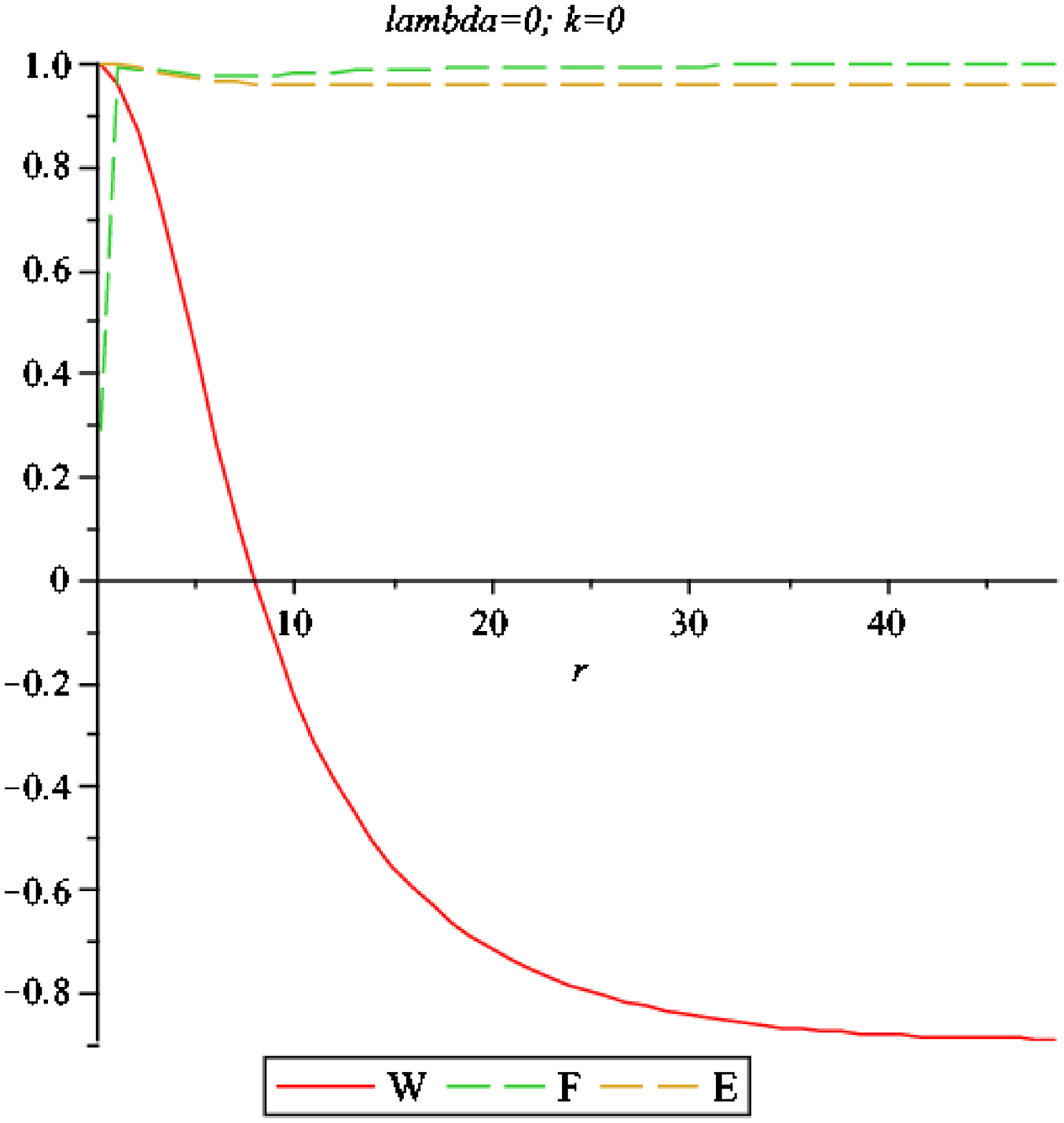} 
\hfill\epsfysize=45mm\epsfbox{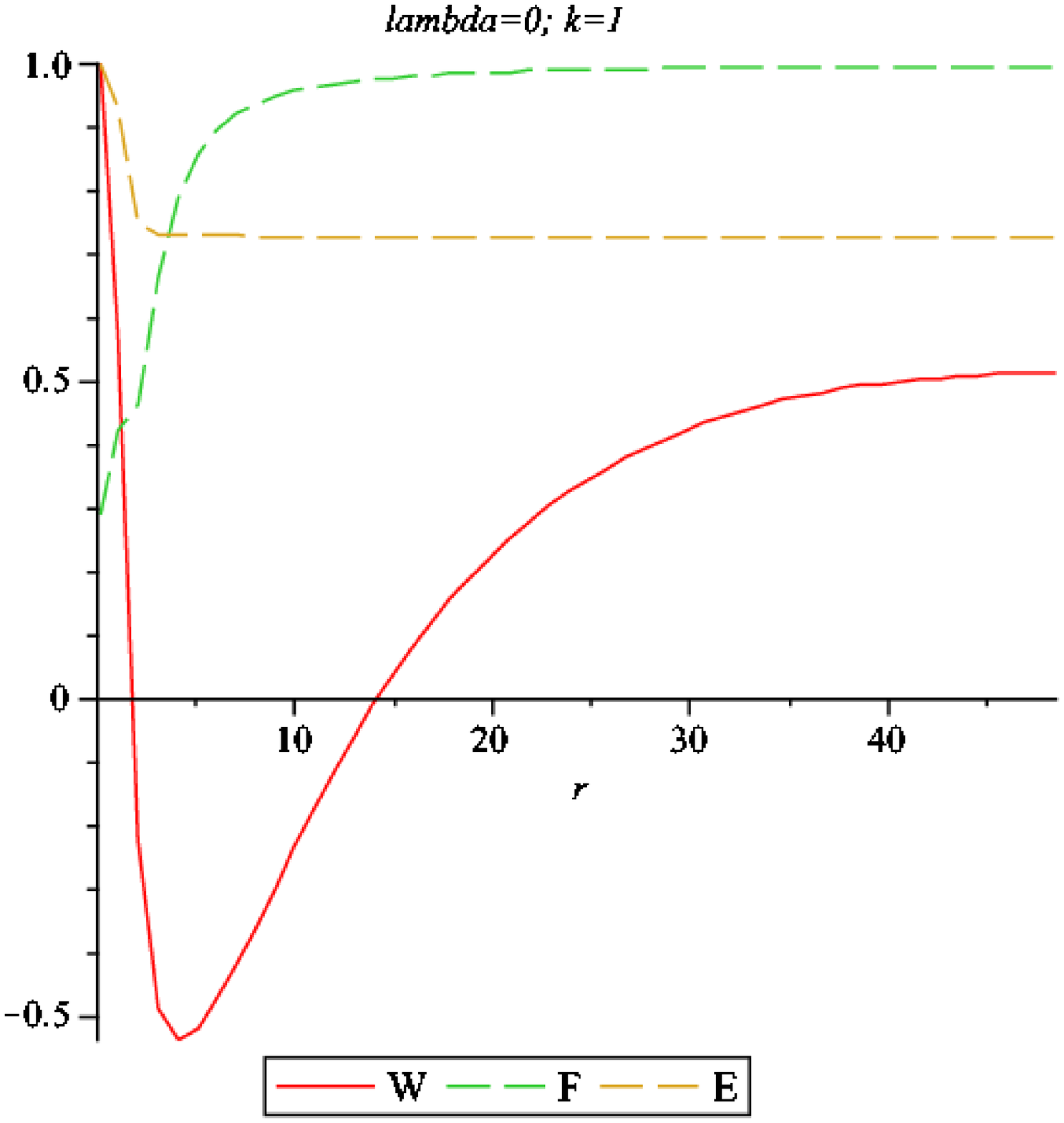}
\hfill}
Fig.1 Solution for $\Lambda=0$ for $\kappa=0$ and $\kappa=1$ respectively.
\\

\centerline{
\hfill\epsfysize=45mm\epsfbox{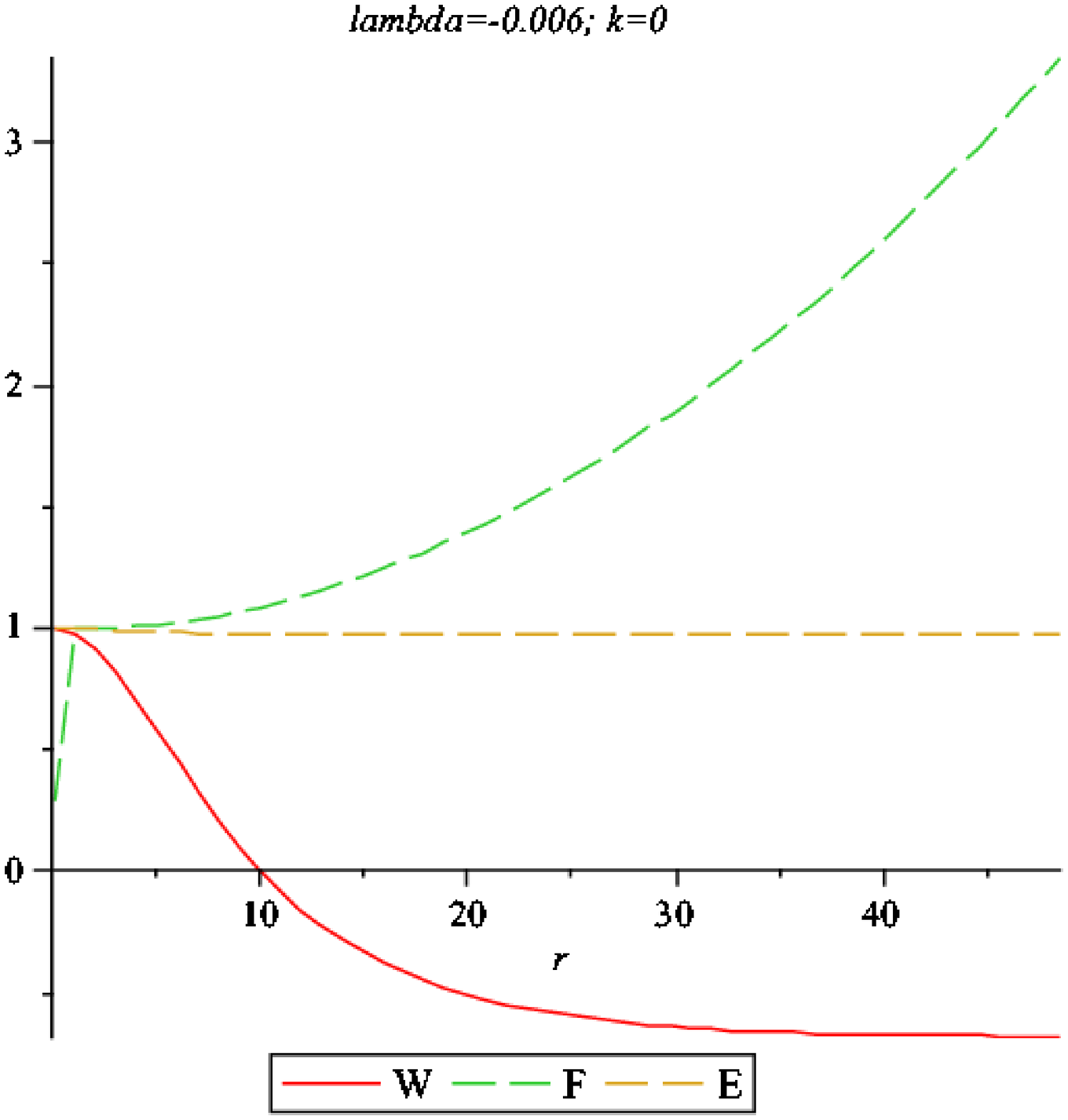}
\hfill\epsfysize=45mm\epsfbox{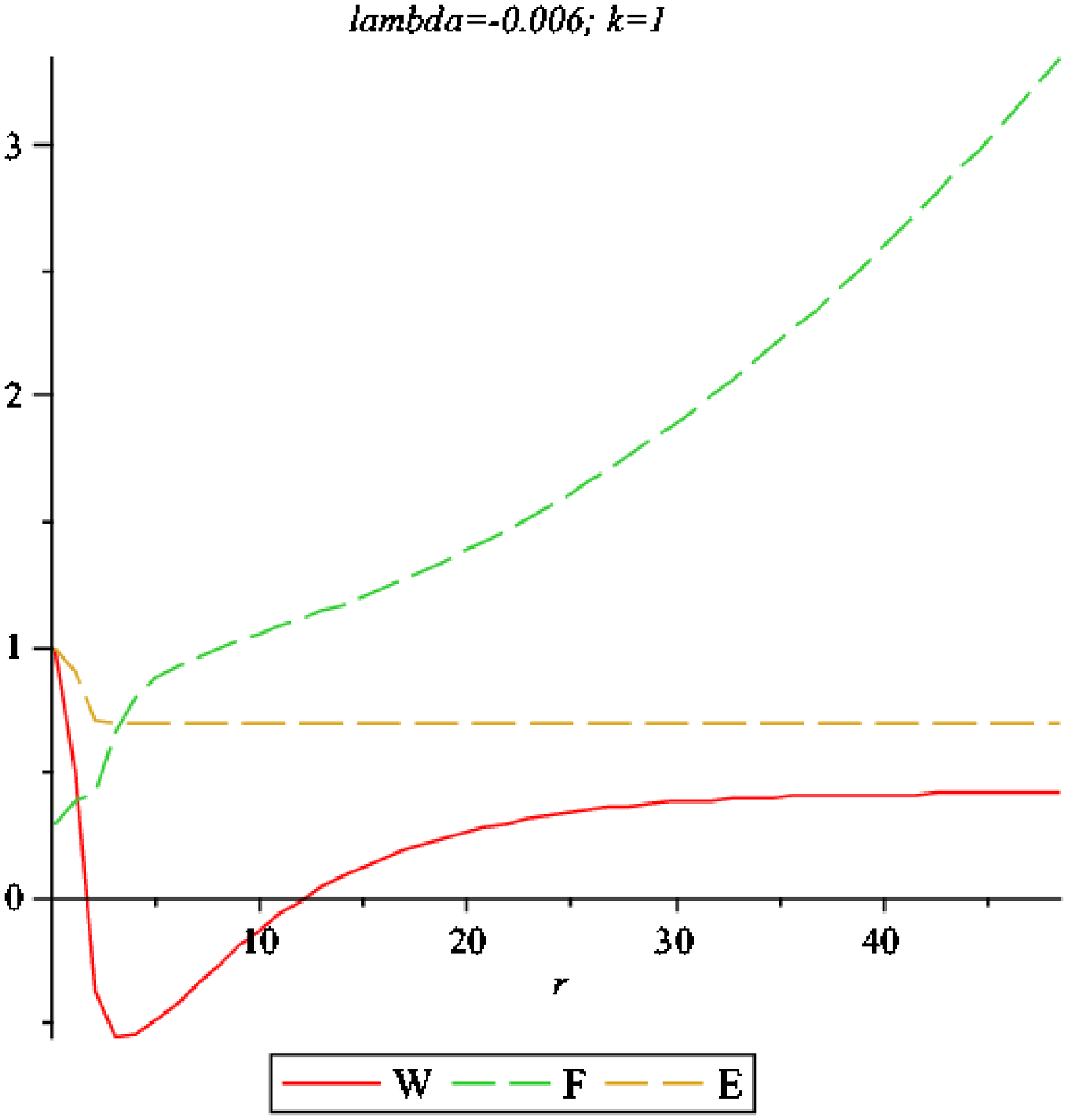}
\hfill}
Fig. 2 Solution for $\Lambda=-0.006$ for $\kappa=0$ and $\kappa=1$ respectively.
\\

\centerline{
\hfill\epsfysize=45mm\epsfbox{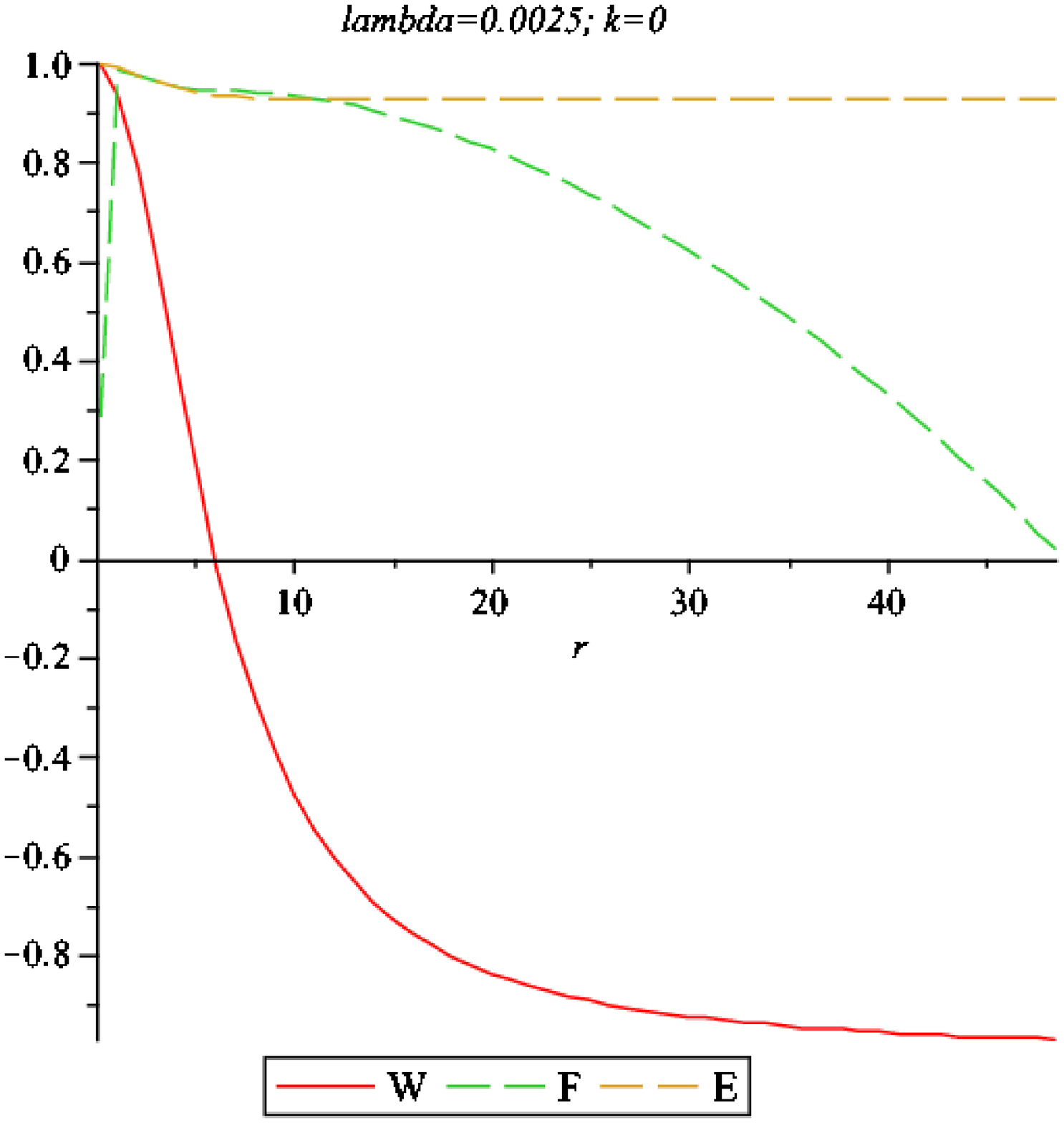}
\hfill\epsfysize=45mm\epsfbox{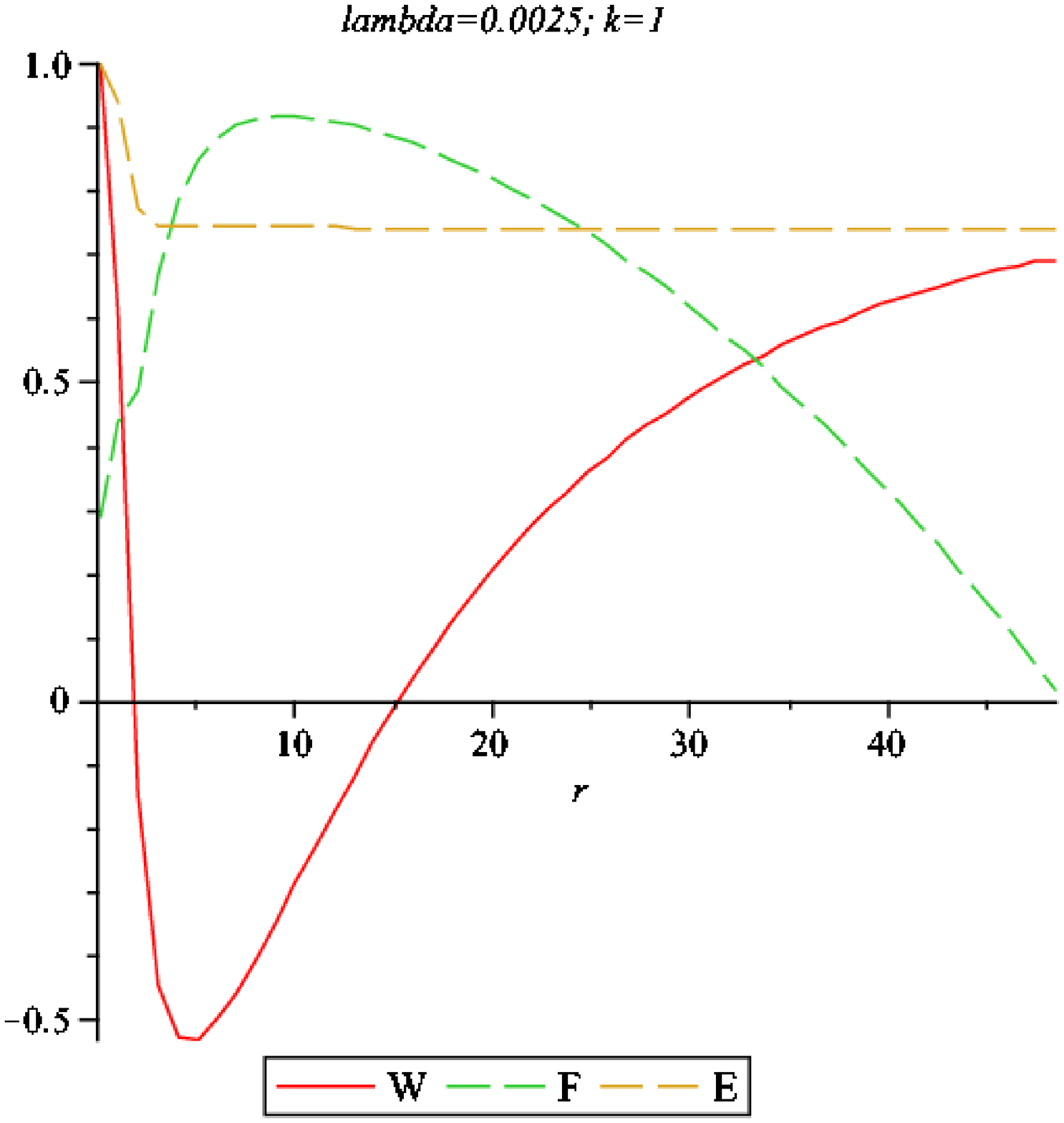}
\hfill}
Fig. 3 Solution for $\Lambda=0.0025$ for $\kappa=0$ and $\kappa=1$ respectively.
\\
From our numerical solutions, we see that the GB term increases the number of nodes of the Yang-Mills field.
Further, we see that for positive $\Lambda$ the solution develops  a singularity, while for negative $\Lambda$ it remains singular free.

A matter field term in the action will lead to an extra term inside the square root of Eq.(16), for example in the case of the 
5-dimensional  Einstein-Maxwell-GB model \cite{Thib} and the 5-dimensional Einstein-Yang-Mills-GB model with the Wu-Yang ansatz \cite{Hab}.
In these models, however, there are no additional equations for the Maxwell field and YM field respectively. So an analytic solution for $F(r)$ 
is obtained.
When one simultaneously tries to solve the Einstein equation and matter field equations, then it is not easy to obtain an analytic 
expression for $F(r)$, as is the case of our EYM-GB model.

However we can  analyse the equation for $F(r)$ when $W$ becomes a  constant :
\begin{equation}
F_r=\frac{2r(1-F)-\frac{2\Lambda r^3}{3}}{r^2+4\kappa (1-F)}.
\end{equation}

The solution is
\begin{equation}
F(r)=1+\frac{r^2}{4\kappa}\pm \frac{1}{12\kappa}\sqrt{(9+12\kappa\Lambda)r^4+144\kappa^2+24\kappa M},
\end{equation}
with $M$ an integration constant. 
Since horizons occur where $F(r)=0$, we can
expect cosmological- and event horizons. One can easily check that the zero's  of $F(r)$ are 
\begin{equation}
r_h=\pm\Bigl(\frac{3\pm \sqrt{9-2\Lambda M}}{\Lambda}\Bigr)^{\frac{1}{2}}.
\end{equation}
So the horizon radius depends only on suitable combinations of $\Lambda$ and $M$. 
The expression inside the square root becomes negative (and hence $F(r)$ is singular) for 
\begin{equation}
r_s\leq\Bigl(\frac{-8\kappa(6\kappa+M)}{4\kappa\Lambda +3}\Bigr)^{\frac{1}{4}}.
\end{equation}
Depending on the parameters, this singular surface can be shielded by the event horizon (otherwise, it will be naked).
This is well known behaviour in the models where the equation for $F(r)$ decouples from the matter field equation.
One should like to prove that for negative $\Lambda$ that F(r) has no zero's and is regular everywhere in our model. 
This is currently under study.

\section{ Conclusion and outlook}
A 5-dimensional spherically symmetric particle-like solution is found in the Einstein-Yang-Mills Gauss-Bonnet
model. As in other studies in higher dimensional cosmological models, a negative cosmological constant seems
to favor for stability and results in most cases in asymptotically anti de Sitter space time.

In our 5-dimensional EYM-GB model, we also find a profound  influence of a negative cosmological constant on the behaviour 
of horizons. The appearance of horizons in E-GB models is not surprising. These GB black holes are found by 
many authors. However, the lacking of horizons in the EYM-GB model  for suitable negative cosmological constant is quite new.
The explanation for this behaviour  must come from the YM term on the right hand side of Eq.(9). 
The zero's of $F(r)$ will depend on the behaviour $W(r)$.

There could be a connection of the solution presented here with the AdS/CFT correspondence. As mentioned before, no analytic
expression for $F(r)$ available. Moreover, to obtain the (n-1)-dimensional entropy on the brane, one needs the junction conditions
at the brane (\cite{Ver,Og}), which becomes very complicated in the EYM-GB model. The junction condition also introduces a brane tension. 
This tension must cancel the cosmological constant, in order to obtain the desired CFT correspondence. 
The contribution of YM field on the junction could have profound impact on the tension of the brane and the role of a cosmological 
constant could be different.
So the strong influence of a small cosmological constant on the eventually formed GB black hole in our model is quite clear from the 
consideration mentioned above.


\end{document}